\documentclass[aps,prl,twocolumn,showpacs,superscriptaddress]{revtex4-1}
\usepackage{graphicx} % Include figure files
\usepackage{dcolumn}  % Align table columns on decimal point
\usepackage{amsmath}
\usepackage{epsfig}
\usepackage{multirow}
\usepackage{mciteplus}

\long\def\inst#1{\par\nobreak\kern 4pt\nobreak
    {\itshape #1}\par\vskip 10pt plus 3pt minus 3pt}

\graphicspath{{./figs/}}
\DeclareGraphicsExtensions{.pdf,.PDF,png,.PNG}

\usepackage{ifthen}  % for conditional statements
\newboolean{uprightparticles}
\setboolean{uprightparticles}{false} % Set true for upright particle symbols

\usepackage{xspace} 
\usepackage{upgreek}

\usepackage{relsize}
\def\babar{\mbox{\slshape B\kern-0.1em{\smaller A}\kern-0.1em
    B\kern-0.1em{\smaller A\kern-0.2em R}}\xspace}

\def\argus  {\mbox{ARGUS}\xspace}

\def\pep    {PEP-II}

\def\MagUp {\mbox{\em Mag\kern -0.05em Up}\xspace}

\ifthenelse{\boolean{uprightparticles}}%
{

 \def\Pmu         {\ensuremath{\upmu}\xspace}

 \def\Ppi         {\ensuremath{\uppi}\xspace}

 \def\PDelta      {\ensuremath{\Delta}\xspace}                 
 \def\PXi      {\ensuremath{\Xi}\xspace}                 
 \def\PLambda      {\ensuremath{\Lambda}\xspace}                 
 \def\PSigma      {\ensuremath{\Sigma}\xspace}                 
 \def\POmega      {\ensuremath{\Omega}\xspace}                 
 \def\PUpsilon      {\ensuremath{\Upsilon}\xspace}                 
 
 \def\PB      {\ensuremath{\mathrm{B}}\xspace}                 
                  
 \def\PD      {\ensuremath{\mathrm{D}}\xspace}

 \def\PK      {\ensuremath{\mathrm{K}}\xspace}

 \def\Pc      {\ensuremath{\mathrm{c}}\xspace}                 
                  
 \def\Pe      {\ensuremath{\mathrm{e}}\xspace}

 \def\Pi      {\ensuremath{\mathrm{i}}\xspace}

 \def\Pq      {\ensuremath{\mathrm{q}}\xspace}

}
{

 \def\Pmu         {\ensuremath{\mu}\xspace}

 \def\Ppi         {\ensuremath{\pi}\xspace}

 \mathchardef\PDelta="7101
 \mathchardef\PXi="7104
 \mathchardef\PLambda="7103
 \mathchardef\PSigma="7106
 \mathchardef\POmega="710A
 \mathchardef\PUpsilon="7107
                  
 \def\PB      {\ensuremath{B}\xspace}                 
                  
 \def\PD      {\ensuremath{D}\xspace}

 \def\PK      {\ensuremath{K}\xspace}

 \def\Pc      {\ensuremath{c}\xspace}                 
                  
 \def\Pe      {\ensuremath{e}\xspace}

 \def\Pi      {\ensuremath{i}\xspace}

 \def\Pq      {\ensuremath{q}\xspace}

}

\makeatletter
\@ifundefined{ptsize}{
\newcommand{\miniscule}{\@setfontsize\miniscule{5}{6}}% assumes 11pt
}
{
\ifcase \@ptsize \relax% 10pt
  \newcommand{\miniscule}{\@setfontsize\miniscule{4}{5}}% \tiny: 5/6
\or% 11pt
  \newcommand{\miniscule}{\@setfontsize\miniscule{5}{6}}% \tiny: 6/7
\or% 12pt
  \newcommand{\miniscule}{\@setfontsize\miniscule{5}{6}}% \tiny: 6/7
\fi
}
\makeatother

\DeclareRobustCommand{\optbar}[1]{\shortstack{{\miniscule (\rule[.5ex]{1.25em}{.18mm})}
  \\ [-.7ex] $#1$}}

\def\ep         {{\ensuremath{\Pe^+}}\xspace}
\def\epm        {{\ensuremath{\Pe^\pm}}\xspace} 
\def\epem       {{\ensuremath{\Pe^+\Pe^-}}\xspace}

\def\mup        {{\ensuremath{\Pmu^+}}\xspace}
\def\mumu       {{\ensuremath{\Pmu^+\Pmu^-}}\xspace}

\def\ellm       {{\ensuremath{\ell^-}}\xspace}
\def\ellp       {{\ensuremath{\ell^+}}\xspace}

\def\quark     {{\ensuremath{\Pq}}\xspace}
\def\quarkbar  {{\ensuremath{\overline \quark}}\xspace}
\def\qqbar     {{\ensuremath{\quark\quarkbar}}\xspace}

\def\cquark    {{\ensuremath{\Pc}}\xspace}
\def\cquarkbar {{\ensuremath{\overline \cquark}}\xspace}
\def\ccbar     {{\ensuremath{\cquark\cquarkbar}}\xspace}

\def\pion   {{\ensuremath{\Ppi}}\xspace}

\def\pip    {{\ensuremath{\pion^+}}\xspace}
\def\pim    {{\ensuremath{\pion^-}}\xspace}

\def\kaon    {{\ensuremath{\PK}}\xspace}
  \def\Kbar    {{\kern 0.2em\overline{\kern -0.2em \PK}{}}\xspace}

\def\KorKbar    {\kern 0.18em\optbar{\kern -0.18em K}{}\xspace}

\def\Kp      {{\ensuremath{\kaon^+}}\xspace}
\def\Km      {{\ensuremath{\kaon^-}}\xspace}

\def\KpKm     {\ensuremath{\Kp \kern -0.16em \Km}\xspace}

  \def\Dbar    {{\kern 0.2em\overline{\kern -0.2em \PD}{}}\xspace}
\def\D       {{\ensuremath{\PD}}\xspace}

\def\DorDbar    {\kern 0.18em\optbar{\kern -0.18em D}{}\xspace}
\def\Dz      {{\ensuremath{\D^0}}\xspace}

\def\Dstarp  {{\ensuremath{\D^{*+}}}\xspace}

\def\B       {{\ensuremath{\PB}}\xspace}
\def\Bbar    {{\ensuremath{\kern 0.18em\overline{\kern -0.18em \PB}{}}}\xspace}

\def\BorBbar    {\kern 0.18em\optbar{\kern -0.18em B}{}\xspace}
\def\Bz      {{\ensuremath{\B^0}}\xspace}
\def\Bzb     {{\ensuremath{\Bbar{}^0}}\xspace}
\def\Bu      {{\ensuremath{\B^+}}\xspace}
\def\Bub     {{\ensuremath{\B^-}}\xspace}

\def\BpBm    {\ensuremath{\Bu {\kern -0.16em \Bub}}\xspace}
\def\BzBzb   {\ensuremath{\Bz {\kern -0.16em \Bzb}}\xspace}
\def\BB      {\ensuremath{\B\Bbar}\xspace}

  \def\Y#1S{\ensuremath{\PUpsilon{(#1S)}}\xspace}% no space before {...}!

\def\FourS {{\Y4S}}

\def\Lbar        {{\ensuremath{\kern 0.1em\overline{\kern -0.1em\PLambda}}}\xspace}
\def\LorLbar    {\kern 0.18em\optbar{\kern -0.18em \PLambda}{}\xspace}

\def\BF         {{\ensuremath{\mathcal{B}}}\xspace}

\def\BR         {\BF}

\def\to                 {\ensuremath{\rightarrow}\xspace}

\def\order   {{\ensuremath{\mathcal{O}}}\xspace}

\newcommand{\dm}{{\ensuremath{\Delta m}}\xspace}

\def\AT#1     {\ensuremath{A_{\mathrm{T}}^{#1}}\xspace}           % 2

\def\C#1      {\ensuremath{\mathcal{C}_{#1}}\xspace}                       % 9
\def\Cp#1     {\ensuremath{\mathcal{C}_{#1}^{'}}\xspace}                    % 7
\def\Ceff#1   {\ensuremath{\mathcal{C}_{#1}^{\mathrm{(eff)}}}\xspace}        % 9  
\def\Cpeff#1  {\ensuremath{\mathcal{C}_{#1}^{'\mathrm{(eff)}}}\xspace}       % 7
\def\Ope#1    {\ensuremath{\mathcal{O}_{#1}}\xspace}                       % 2
\def\Opep#1   {\ensuremath{\mathcal{O}_{#1}^{'}}\xspace}                    % 7

\newcommand{\tev}{\ifthenelse{\boolean{inbibliography}}{\ensuremath{~T\kern -0.05em eV}\xspace}{\ensuremath{\mathrm{\,Te\kern -0.1em V}}}\xspace}
\newcommand{\gev}{\ensuremath{\mathrm{\,Ge\kern -0.1em V}}\xspace}
\newcommand{\mev}{\ensuremath{\mathrm{\,Me\kern -0.1em V}}\xspace}
\newcommand{\kev}{\ensuremath{\mathrm{\,ke\kern -0.1em V}}\xspace}
\newcommand{\ev}{\ensuremath{\mathrm{\,e\kern -0.1em V}}\xspace}
\newcommand{\gevc}{\ensuremath{{\mathrm{\,Ge\kern -0.1em V\!/}c}}\xspace}
\newcommand{\mevc}{\ensuremath{{\mathrm{\,Me\kern -0.1em V\!/}c}}\xspace}
\newcommand{\gevcc}{\ensuremath{{\mathrm{\,Ge\kern -0.1em V\!/}c^2}}\xspace}
\newcommand{\gevgevcccc}{\ensuremath{{\mathrm{\,Ge\kern -0.1em V^2\!/}c^4}}\xspace}
\newcommand{\mevcc}{\ensuremath{{\mathrm{\,Me\kern -0.1em V\!/}c^2}}\xspace}

\def\invfb   {\ensuremath{\mbox{\,fb}^{-1}}\xspace}

\def\order{{\ensuremath{\mathcal{O}}}\xspace}
\newcommand{\chisq}{\ensuremath{\chi^2}\xspace}

\def\gsim{{~\raise.15em\hbox{$>$}\kern-.85em
          \lower.35em\hbox{$\sim$}~}\xspace}
\def\lsim{{~\raise.15em\hbox{$<$}\kern-.85em
          \lower.35em\hbox{$\sim$}~}\xspace}

\newcommand{\lum} {\ensuremath{\mathcal{L}}\xspace}
\def\evtgen     {\mbox{\textsc{EvtGen}}\xspace}
\def\geantfour  {\mbox{\tt GEANT 4}\xspace} % same as BaBar format

\def\photos     {\mbox{\textsc{Photos}}\xspace}
\def\tell1  {TELL1\xspace}
\def\ukl1   {UKL1\xspace}

\def\usedlumi   {\ensuremath{39.3\pm0.2}}  % what was actually used
\def\totallumiNoerr  {\ensuremath{468.2}}
\def\totallumi  {\ensuremath{\totallumiNoerr\pm2.0}}

\def\nsig       {\ensuremath{N_{\rm sig}}}
\def\nnorm      {\ensuremath{N_{\rm norm}}}

\def\DzToKPiee   {\ensuremath{\Dz\to\Km\pip\epem}}

\def\KKPiPi      {\ensuremath{\Km\Kp\pip\pim}}
\def\DzToKKPiPi  {\ensuremath{\Dz\to\KKPiPi}}

\def\KPiPiPi      {\ensuremath{\Km\pip\pip\pim}}
\def\DzToKPiPiPi  {\ensuremath{\Dz\to\KPiPiPi}}

\def\PiPiPiPi      {\ensuremath{\pim\pip\pip\pim}}
\def\DzToPiPiPiPi  {\ensuremath{\Dz\to\PiPiPiPi}}

\def\DzToKPi     {\ensuremath{\Dz\to\Km\pip}}

\def\multiKKPiPi      {\ensuremath{2.4}}
\def\multiPiPiPiPi    {\ensuremath{4.4}}

\def\BFDzToKPiPiPi  {\ensuremath{3.98\pm0.08\pm0.10}} % x 10-2
\def\BFDzToKPi      {\ensuremath{3.89\pm0.04}} % PDG

\def\effKPiNoErr       {\ensuremath{27.4}}
\def\effKKPiPiNoErr    {\ensuremath{19.2}}
\def\effKPiPiPiNoErr   {\ensuremath{20.1}}
\def\effPiPiPiPiNoErr  {\ensuremath{24.7}}

\def\effKPi       {\ensuremath{\effKPiNoErr\pm0.2}}
\def\effKKPiPi    {\ensuremath{\effKKPiPiNoErr\pm0.2}}
\def\effKPiPiPi   {\ensuremath{\effKPiPiPiNoErr\pm0.2}}
\def\effPiPiPiPi  {\ensuremath{\effPiPiPiPiNoErr\pm0.2}}

\def\yieldKPi       {\ensuremath{1\,881\,950\pm1380}} % rounded
\def\yieldKKPiPi    {\ensuremath{8480\pm110}}       % rounded
\def\yieldKPiPiPi   {\ensuremath{260\,870\pm520}}     % rounded
\def\yieldPiPiPiPi  {\ensuremath{28\,470\pm220}}      % rounded

\def\systPiPiPiPi    {\ensuremath{6.8}}
\def\systKPiPiPi     {\ensuremath{4.7}}
\def\systKKPiPi      {\ensuremath{6.6}}

\def\DzTohhmumuOS  {\ensuremath{\Dz\to h^{\prime -}h^{+}\mumu}}

\def\DzTohhllSS  {\ensuremath{\Dz\to h^{\prime -} h^{-}\ell^{\prime +} \ellp}}

\def\DzTohhllOS    {\ensuremath{\Dz\to h^{\prime -} h^{+}\ell^{\prime\pm} \ell^{\mp}}}

\def\effKPieeSSNoErr     {\ensuremath{3.19}}
\def\effKPimumuSSNoErr   {\ensuremath{3.30}}
\def\effKPiemuSSNoErr    {\ensuremath{3.48}}

\def\effKKeeSSNoErr      {\ensuremath{3.25}}
\def\effKKmumuSSNoErr    {\ensuremath{6.21}}
\def\effKKemuSSNoErr     {\ensuremath{4.63}}

\def\effPiPieeSSNoErr    {\ensuremath{4.38}}
\def\effPiPimumuSSNoErr  {\ensuremath{4.91}}
\def\effPiPiemuSSNoErr   {\ensuremath{4.38}}

\def\effKPiemuOSNoErr    {\ensuremath{3.65}}
\def\effKKemuOSNoErr     {\ensuremath{4.83}}
\def\effPiPiemuOSNoErr   {\ensuremath{4.79}}

\def\effKPieeSS     {\ensuremath{\effKPieeSSNoErr\pm0.05}}
\def\effKPimumuSS   {\ensuremath{\effKPimumuSSNoErr\pm0.05}}
\def\effKPiemuSS    {\ensuremath{\effKPiemuSSNoErr\pm0.04}}

\def\effKKeeSS      {\ensuremath{\effKKeeSSNoErr\pm0.04}}
\def\effKKmumuSS    {\ensuremath{\effKKmumuSSNoErr\pm0.06}}
\def\effKKemuSS     {\ensuremath{\effKKemuSSNoErr\pm0.05}}

\def\effPiPieeSS    {\ensuremath{\effPiPieeSSNoErr\pm0.05}}
\def\effPiPimumuSS  {\ensuremath{\effPiPimumuSSNoErr\pm0.05}}
\def\effPiPiemuSS   {\ensuremath{\effPiPiemuSSNoErr\pm0.05}}

\def\effKPiemuOS    {\ensuremath{\effKPiemuOSNoErr\pm0.05}}
\def\effPiPiemuOS   {\ensuremath{\effPiPiemuOSNoErr\pm0.06}}
\def\effKKemuOS     {\ensuremath{\effKKemuOSNoErr\pm0.05}}

\def\multiKKPiPi      {\ensuremath{2.4}}
\def\multiPiPiPiPi    {\ensuremath{4.4}}

\def\multiPiPiemuOS   {\ensuremath{4.5}}
\def\multiKKemuOS     {\ensuremath{7.1}}

\def\obsKPieeSS     {\ensuremath{-0.23\pm0.97\pm1.28}} 
\def\obsKPimumuSS   {\ensuremath{-0.03\pm2.10\pm0.40}}
\def\obsKPiemuSS    {\ensuremath{3.87\pm3.96\pm2.36}}

\def\obsKKeeSS      {\ensuremath{0.30\pm1.08\pm0.41}}
\def\obsKKmumuSS    {\ensuremath{-1.09\pm1.29\pm0.42}}
\def\obsKKemuSS     {\ensuremath{1.93\pm1.92\pm0.83}}
\def\obsKKemuOS     {\ensuremath{4.09\pm3.00\pm1.59}}

\def\obsPiPieeSS    {\ensuremath{0.22\pm3.15\pm0.54}}
\def\obsPiPimumuSS  {\ensuremath{6.69\pm4.88\pm0.80}}
\def\obsPiPiemuSS   {\ensuremath{12.42\pm5.30\pm1.45}}

\def\obsKPiemuOS    {\ensuremath{2.52\pm4.60\pm1.35}}
\def\obsPiPiemuOS   {\ensuremath{1.37\pm6.15\pm1.28}}

\def\ulactKPieeSS     {\ensuremath{5.0}}
\def\ulactKPimumuSS   {\ensuremath{5.3}}
\def\ulactKPiemuSS    {\ensuremath{21.0}}

\def\ulactPiPieeSS    {\ensuremath{9.1}}
\def\ulactPiPimumuSS  {\ensuremath{15.2}}
\def\ulactPiPiemuSS   {\ensuremath{30.6}}

\def\ulactKPiemuOS    {\ensuremath{19.0}}
\def\ulactPiPiemuOS   {\ensuremath{17.1}}

\def\ulactKKeeSS      {\ensuremath{3.4}}
\def\ulactKKmumuSS    {\ensuremath{1.0}}
\def\ulactKKemuSS     {\ensuremath{5.8}}
\def\ulactKKemuOS     {\ensuremath{10.0}}

\def\systPiPiPiPi    {\ensuremath{6.8}}
\def\systKPiPiPi     {\ensuremath{4.7}}
\def\systKKPiPi      {\ensuremath{6.6}}

\def\systBFKPieeSS     {\ensuremath{2.11}}
\def\systBFKPimumuSS   {\ensuremath{0.64}}
\def\systBFKPiemuSS    {\ensuremath{3.56}}

\def\systBFPiPieeSS    {\ensuremath{0.67}}
\def\systBFPiPimumuSS  {\ensuremath{0.91}}
\def\systBFPiPiemuSS   {\ensuremath{1.85}}

\def\systBFKPiemuOS    {\ensuremath{1.95}}
\def\systBFPiPiemuOS   {\ensuremath{1.45}}

\def\systBFKKeeSS      {\ensuremath{0.58}} % line 106
\def\systBFKKmumuSS    {\ensuremath{0.32}}
\def\systBFKKemuSS     {\ensuremath{0.84}}
\def\systBFKKemuOS     {\ensuremath{1.45}}

\def\bfactKPieeSS     {\ensuremath{-0.38\pm1.60\pm\systBFKPieeSS}}
\def\bfactKPimumuSS   {\ensuremath{-0.05\pm3.34\pm\systBFKPimumuSS}}
\def\bfactKPiemuSS    {\ensuremath{5.84\pm5.97\pm\systBFKPiemuSS}}

\def\bfactPiPieeSS    {\ensuremath{0.27\pm3.90\pm\systBFPiPieeSS}}
\def\bfactPiPimumuSS  {\ensuremath{7.40\pm5.40\pm\systBFPiPimumuSS}}
\def\bfactPiPiemuSS   {\ensuremath{15.41\pm6.59\pm\systBFPiPiemuSS}}

\def\bfactKPiemuOS    {\ensuremath{3.62\pm6.61\pm\systBFKPiemuOS}}
\def\bfactPiPiemuOS   {\ensuremath{1.55\pm6.97\pm\systBFPiPiemuOS}}

\def\bfactKKeeSS      {\ensuremath{0.43\pm1.54\pm\systBFKKeeSS}}
\def\bfactKKmumuSS    {\ensuremath{-0.81\pm0.96\pm\systBFKKmumuSS}}
\def\bfactKKemuSS     {\ensuremath{1.93\pm1.93\pm\systBFKKemuSS}}
\def\bfactKKemuOS     {\ensuremath{3.93\pm2.89\pm\systBFKKemuOS}}

\newcommand{\BABARPubYear}    {19}
\newcommand{\BABARPubNumber}  {002}
\newcommand{\SLACPubNumber} {17424}

\begin{document}

\begin{@twocolumnfalse}
\noindent \babar-PUB-\BABARPubYear/\BABARPubNumber\\
\noindent SLAC-PUB-\SLACPubNumber\\[10mm]
\end{@twocolumnfalse}

\title{{\Large \bf \boldmath Search for Rare or Forbidden Decays of the $D^{0}$ Meson}}

\author{J.~P.~Lees}
\author{V.~Poireau}
\author{V.~Tisserand}
\affiliation{Laboratoire d'Annecy-le-Vieux de Physique des Particules (LAPP), Universit\'e de Savoie, CNRS/IN2P3,  F-74941 Annecy-Le-Vieux, France}
\author{E.~Grauges}
\affiliation{Universitat de Barcelona, Facultat de Fisica, Departament ECM, E-08028 Barcelona, Spain }
\author{A.~Palano}
\affiliation{INFN Sezione di Bari and Dipartimento di Fisica, Universit\`a di Bari, I-70126 Bari, Italy }
\author{G.~Eigen}
\affiliation{University of Bergen, Institute of Physics, N-5007 Bergen, Norway }
\author{D.~N.~Brown}
\author{Yu.~G.~Kolomensky}
\affiliation{Lawrence Berkeley National Laboratory and University of California, Berkeley, California 94720, USA }
\author{M.~Fritsch}
\author{H.~Koch}
\author{T.~Schroeder}
\affiliation{Ruhr Universit\"at Bochum, Institut f\"ur Experimentalphysik 1, D-44780 Bochum, Germany }
% ...
\collaboration{The \babar\ Collaboration}
\noaffiliation

\begin{abstract}
\noindent We present a search for nine lepton-number-violating and
three lepton-flavor-violating neutral charm decays of the type
\DzTohhllSS\ and \DzTohhllOS, where $h$ and $h^{\prime}$ represent a $K$
or $\pi$ meson and $\ell$ and $\ell^{\prime}$ an electron or muon. The
analysis is based on $468$\invfb\ of $\epem$ annihilation data
collected at or close to the \FourS\ resonance with the
\babar\ detector at the SLAC National Accelerator Laboratory. No
significant signal is observed for any of the twelve modes, and we
establish 90\% confidence level upper limits on the branching
fractions in the range $(\ulactKKmumuSS - \ulactPiPiemuSS)\times
10^{-7}$. The limits are between 1 and 3 orders of magnitude
more stringent than previous measurements.
\end{abstract}

  
\maketitle

Lepton-flavor-violating and lepton-number-violating neutral charm
decays can be used to investigate physics beyond the standard model
(SM) of particle physics. A potential set of decays for study are of
the form \DzTohhllSS\ and \DzTohhllOS, where $h$ and $h^{\prime}$
represent a $K$ or $\pi$ meson and $\ell$ and $\ell^{\prime}$ an
electron or muon~\cite{conjugate}.

The \DzTohhllOS\ decay modes with two opposite-charge,
different-flavor leptons in the final state are
lepton-flavor-violating (LFV). They are essentially prohibited in the
SM because they can occur only through lepton  
mixing~\cite{GUADAGNOLI201554}. The \DzTohhllSS\ decay modes with two
same-charge leptons are both lepton-flavor violating and
lepton-number violating (LNV) and are forbidden in the SM in
low-energy collisions or decays. However, LNV processes can occur in
extremely high-energy or high-density
interactions~\cite{Klinkhamer:1984di}.

Lepton-number violation is a necessary condition for leptogenesis as
an explanation of the baryon asymmetry of the
Universe~\cite{DAVIDSON2008105}. If neutrinos are Majorana fermions,
the neutrino and antineutrino are the same particle, and some LNV
processes become possible~\cite{Majorana:1937vz}.  Many models beyond
the SM allow lepton-number violation. Most models have made
predictions for, or used constraints from, three-body decays of the form
$\D\to Ml'l$ or $\B\to Ml'l$, where $M$ is a
meson~\cite{Paul:2011ar,Paul:2012ab,Burdman:2001tf,Fajfer:2005ke,PhysRevD.76.074010,Atre:2009aa,Yuan:2013yba}. However,
some models that consider LFV and LNV four-body charm decays predict
branching fractions up to $\order(10^{-6})$ to $\order(10^{-5})$,
approaching those accessible with current
data~\cite{Atre:2009aa,Yuan:2013yba,1674-1137-39-1-013101}.

The branching fractions $\BR(\DzTohhmumuOS)$ and $\BR(\DzToKPiee)$
have recently been determined to be $\order(10^{-7})$ to
$\order(10^{-6})$~\cite{LHCb-PAPER-2015-043,LHCb-PAPER-2017-019,Lees:2018zz},
compatible with SM predictions~\cite{Cappiello:2012vg,Gudrun:035041}. The most
stringent existing upper limits on the branching fractions for the LFV
and LNV four-body decays of the type \DzTohhllOS\ and \DzTohhllSS\ are
in the range $(0.3-55.3)\times 10^{-5}$ at the 90\% confidence level
(C.L.)~\cite{Aitala:2000kk,PDG2018,Ablikim:2019gvd}. For the LFV decays $\Dz\to
V\ell^{\prime +}\ellm$, where $V$ is an intermediate resonance such as
a $\rho$ or $\phi$ meson decaying to $h^{\prime -} h^{+}$, the 90\%
C.L. limits are in the range $(3.4-118)\times
10^{-5}$~\cite{Freyberger:1996it,Aitala:2000kk,PDG2018}. Searches for
Majorana neutrinos in $D^{+}_{(s)}\to\pim\mup\mup$ decays have placed upper
limits on the branching fractions as low as $2.2\times 10^{-8}$ at the
90\% C.L.~\cite{2013203}.

In this report we present a search for nine \DzTohhllSS\ LNV
decays and three \DzTohhllOS\ LFV decays, with data recorded with the
\babar\ detector at the \pep\ asymmetric-energy $\epem$ collider
operated at the SLAC National Accelerator Laboratory. The data sample
corresponds to 424\invfb\ of \epem\ collisions collected at the
center-of-mass (c.m.) energy of the \FourS\ resonance (on peak) and an
additional 44\invfb\ of data collected 40~\mev\ below the
\FourS\ resonance (off peak)~\cite{Lees:2013rw}. The branching
fractions for signal modes with zero, one, or two kaons in the final
state are measured relative to the normalization decays
\DzToPiPiPiPi, \DzToKPiPiPi, and \DzToKKPiPi, respectively. The
\Dz\ mesons are identified from the decay $\Dstarp\to\Dz\pip$
produced in $\epem\to\ccbar$ events.

The \babar\ detector is described in detail in
Refs.~\cite{Aubert:2001tu,TheBABAR:2013jta}. Charged particles are
reconstructed as tracks with a five-layer silicon vertex detector and
a 40-layer drift chamber inside a $1.5\,$T solenoidal magnet. An
electromagnetic calorimeter comprised of 6580 CsI(Tl) crystals is used
to identify electrons and photons. A ring-imaging Cherenkov detector
is used to identify charged hadrons and to provide additional lepton
identification information. Muons are identified with an instrumented
magnetic-flux return.

Monte Carlo (MC) simulation is used to investigate sources of
background contamination, evaluate selection efficiencies, cross-check
the selection procedure, and for studies of systematic effects.  The
signal and normalization channels are simulated with the
\evtgen\ package~\cite{Lange:2001uf}. We generate the signal channel
decays uniformly throughout the four-body phase space, while the
normalization modes include two-body and three-body intermediate
resonances, as well as nonresonant decays. We also generate
$\epem\to\qqbar$ ($q=u,d,s,c$), dimuon, Bhabha elastic
\epem\ scattering, \BB\ background, and two-photon
events~\cite{Ward:2002qq, Sjostrand:1993yb}. Final-state radiation is
generated using \photos~\cite{Golonka:2005pn}. The detector response
is simulated with
\geantfour~\cite{Agostinelli:2002hh,Allison:2006ve}.

In order to optimize the event reconstruction, candidate selection
criteria, multivariate analysis training, and fit procedure, a
rectangular area in the $m(\Dz)$ versus $\dm = m(\Dstarp)-m(\Dz)$
plane is defined, where $m(\Dstarp)$ and $m(\Dz)$ are the
reconstructed masses of the \Dstarp\ and \Dz\ candidates,
respectively. This optimization region is kept hidden (blinded) in
data until the analysis steps are finalized. The blinded region is
approximately 3 times the width of the \dm\ and $m(\Dz)$
resolutions. The \dm\ region is $0.1447<\dm<0.1462\gevcc$ for
all modes. The $m(\Dz)$ signal peak distribution is asymmetric
due to bremsstrahlung emission. The upper $m(\Dz)$
bound on the blinded region is 1.874\gevcc for all modes, and the
lower bound is 1.848\gevcc, 1.852\gevcc, and 1.856\gevcc for modes
with two, one or no electrons, respectively.

Events are required to contain at least five charged tracks. Particle
identification (PID) criteria are applied to all the charged tracks to
identify kaons, pions, electrons, and
muons~\cite{Adam:2004fq,TheBABAR:2013jta}. For modes with two kaons in
the final state, the PID requirement on the kaons is relaxed compared
to the single-kaon modes. This increases the reconstruction efficiency
for the modes with two kaons, with little increase in backgrounds or
misidentified candidates. The PID efficiency depends on the track
momentum, and is in the range $0.96-0.99$ for electrons, $0.60-0.95$
for muons, and $0.90-0.98$ for kaons and pions. The misidentification
probability is typically less than $0.03$ for all selection criteria,
except for the pion selection criteria, where the muon
misidentification rate can be as high as $0.35$ at low momentum.

Candidate \Dz\ mesons are formed from four
charged tracks reconstructed with the appropriate mass hypotheses for
the signal and normalization decays. The four tracks must form a
good-quality vertex with a \chisq\ probability for the vertex fit
greater than 0.005. A bremsstrahlung energy recovery algorithm is
applied to electrons~\cite{Lees:2018zz}. The invariant mass of any \epem\ pair is required
to be greater than $0.1\gevcc$.  For the normalization modes, the
reconstructed \Dz\ meson mass is required to be in the range
$1.81<m(\Dz)<1.91\gevcc$, while for the signal modes, $m(\Dz)$ must be
in the blinded $m(\Dz)$ range defined above.

The candidate \Dstarp\ is formed by combining the \Dz\ candidate with
a charged pion with a momentum in the laboratory frame greater than
$0.1\gevc$. For the normalization mode \DzToKPiPiPi, this pion is
required to have a charge opposite to that of the final-state kaon. A
vertex fit is performed with the \Dz\ mass constrained to its known
value~\cite{PDG2018} and the requirement that the \Dz\ meson and the
pion originate from the \pep\ interaction region. The
\chisq\ probability of the fit is required to be greater than 0.005.
For signal modes with two kaons, the mass difference \dm\ is required
to be $0.141<\dm<0.201\gevcc$. Signal modes with fewer than two kaons
have almost no candidates beyond $\dm=0.149\gevcc$, and the range for
these modes is restricted to $0.141<\dm<0.149\gevcc$.

After the application of the \Dstarp\ vertex fit, the \Dz\ candidate
momentum in the c.m. system, $p^{\ast}$, is required to be greater than
2.4\gevc. This removes most sources of combinatorial background and
also charm hadrons produced in \B\ decays, which are limited to
$p^{\ast} \lesssim 2.2\gevc$~\cite{PhysRevD.69.111104}.

Remaining backgrounds are mainly radiative Bhabha scattering,
initial-state radiation, and two-photon events, which are all rich in
electrons and positrons. We suppress these backgrounds by requiring that the PID
signatures of the hadron candidates be inconsistent with the electron
hypothesis.

Hadronic \Dz\ decays with large branching fractions, where one or
  more charged tracks are misidentified as leptons, will usually have
  reconstructed \Dz\ masses well away from the known
  \Dz\ mass~\cite{PDG2018}. To ensure rejection of this type of
background.  the \Dz\ candidate is also reconstructed assuming the
kaon or pion mass hypothesis for the lepton candidates. If the
resulting \Dz\ candidate mass is within 20\mevcc\ of the known
\Dz\ mass, and if $|\dm|<2\mevcc$, the event is discarded. After these
criteria are applied, the background from these hadronic decays is
negligible.

Two particular sources of background are semileptonic charm decays in
which a charged hadron is misidentified as a lepton; and charm decays
in which the final state contains a neutral particle or more than four
charged tracks. In both cases, tracks can be selected from elsewhere
in the event to form a \Dz\ candidate. To reject these backgrounds, a
multivariate selection based on a Fisher discriminant is
applied~\cite{Fisher:1936et}. The discriminant uses nine input
observables: the momenta of the four tracks used to form the
\Dz\ candidate; the thrust and sphericity of the
\Dstarp\ candidate~\cite{DeRujula:1978vmq}; the angle between the
\Dstarp\ meson candidate sphericity axis and the sphericity axis
defined by the charged particles in the rest of the event (ROE); the
angle between the \Dstarp\ meson candidate thrust axis and the thrust
axis defined by the charged particles in the ROE; and the second
Fox-Wolfram moment~\cite{Fox:1978vw} calculated from the entire event
using both charged and neutral particles. The input observables are
determined in the laboratory frame after the application of the
  \Dstarp\ vertex fit. The discriminant is trained and tested using
MC for the signal modes; for the background, data outside the
optimization region, together with $\epem\to\ccbar$ MC samples, are
used. The training is performed independently for each signal mode. A
requirement on the Fisher discriminant output is chosen such that
approximately 90\% of the simulation signal candidates are
accepted. Depending on the signal mode, this rejects 30\% to 50\% of
the background in data.

The cross feed to one signal mode from the other eleven is estimated
from MC samples to be $\lesssim0.5\%$ in all cases, assuming equal
branching fractions for all signal modes. The cross feed to a specific
normalization mode from the other two normalization modes is predicted
from simulation to be $\lesssim0.7\%$, where the branching fractions
are taken from Ref.~\cite{PDG2018}. Multiple candidates occur in
\multiPiPiemuOS\% to \multiKKemuOS\% of simulated signal events and in
\multiKKPiPi\% to \multiPiPiPiPi\% of the normalization events in
data. If two or more candidates are found in an event, the one with
the highest \Dstarp\ vertex \chisq\ probability is selected. After the
application of all selection criteria and corrections for small
differences between data and MC simulation in tracking and PID
performance derived from high purity control
samples~\cite{TheBABAR:2013jta}, the reconstruction efficiency
$\epsilon_{\rm sig}$ for the simulated signal decays is between 3.2\%
and 6.2\%, depending on the mode. For the normalization decays, the
reconstruction efficiency $\epsilon_{\rm norm}$ is between
\effKKPiPiNoErr\% and \effPiPiPiPiNoErr\%. The difference between
$\epsilon_{\rm sig}$ and $\epsilon_{\rm norm}$ is mainly due to the
momentum dependence of the lepton PID~\cite{TheBABAR:2013jta}.

The signal mode branching fraction $\BR_{\rm sig}$ is determined relative to that of the
normalization decay using
\begin{equation}
\label{eq:ratio}
  \BR_{\rm sig} = \frac{N_{\rm sig}}{N_{\rm norm}} 
  \frac{\epsilon_{\rm norm}}{\epsilon_{\rm sig}} \frac{\lum_{\rm norm}}{\lum_{\rm sig}} \BR_{\rm norm},
\end{equation}

\noindent where \hbox{$\BR_{\rm norm}$} is the
normalization mode branching fraction~\cite{PDG2018}, and $N_{\rm sig}$ and $N_{\rm
  norm}$ are the fitted yields of the signal and normalization mode
decays, respectively. The symbols $\lum_{\rm sig}$ and $\lum_{\rm
  norm}$ represent the integrated luminosities of the data samples
used for the signal (\totallumi\invfb) and the normalization decays
(\usedlumi\invfb), respectively~\cite{Lees:2013rw}. For the signal
modes, we use the on-peak and off-peak data samples, while the
normalization modes use only a subset of the off-peak data.

Each normalization mode yield $N_{\rm norm}$ is extracted by
performing an extended two-dimensional unbinned maximum likelihood
(ML) fit~\cite{Lees:2013gdj} to the observables \dm\ and $m(\Dz)$ in
the range $0.141<\dm<0.149\gevcc$ and $1.81<m(\Dz)<1.91\gevcc$.  The
measured \dm\ and $m(\Dz)$ values are not correlated and are treated
as independent observables in the fits. The probability density
functions (PDFs) in the fits depend on the normalization mode and use
sums of multiple Cruijff~\cite{Lees:2018zz} and Crystal
Ball~\cite{Skwarnicki:1986xj} functions in both \dm\ and $m(\Dz)$. The
functions for each observable use a common mean. The background is
modeled with an \argus\ threshold function~\cite{Albrecht:1990am} for
\dm\ and a Chebyshev polynomial for $m(\Dz)$. The \argus\ end point
parameter is fixed to the kinematic threshold for a
$\Dstarp\to\Dz\pip$ decay. All other PDF parameters, together with the
normalization mode and background yields, are allowed to vary in the
fit. The fitted yields and reconstruction efficiencies for the
normalization modes are given in Table~\ref{tab:norm_fit}.

\begin{table}[htbp!]
\begin{center}
\caption{Summary of fitted candidate yields with statistical
  uncertainties, systematic uncertainties, and reconstruction
  efficiencies for the three normalization modes.}
    \begin{tabular}{lrcc}
      \hline
      \hline \\ [-1em]
        Decay mode  &  \multicolumn{1}{c}{\nnorm}        & Systematic & $\epsilon_{\rm norm}$ \\
            &  \multicolumn{1}{c}{(candidates)}  & (\%)  & (\%)    \\
      \hline \\ [-1em]
      \Dz\to\KPiPiPi  & \yieldKPiPiPi  & \systKPiPiPi & \effKPiPiPi   \\
      \Dz\to\KKPiPi   & \yieldKKPiPi   & \systKKPiPi  & \effKKPiPi       \\
      \Dz\to\PiPiPiPi & \yieldPiPiPiPi & \systPiPiPiPi & \effPiPiPiPi  \\
      \hline
      \hline
    \end{tabular}
    \label{tab:norm_fit}
  \end{center}
\end{table}

\begin{figure}[htbp!]
\begin{center}
  \begin{tabular}{c}
  \includegraphics[width=\columnwidth]{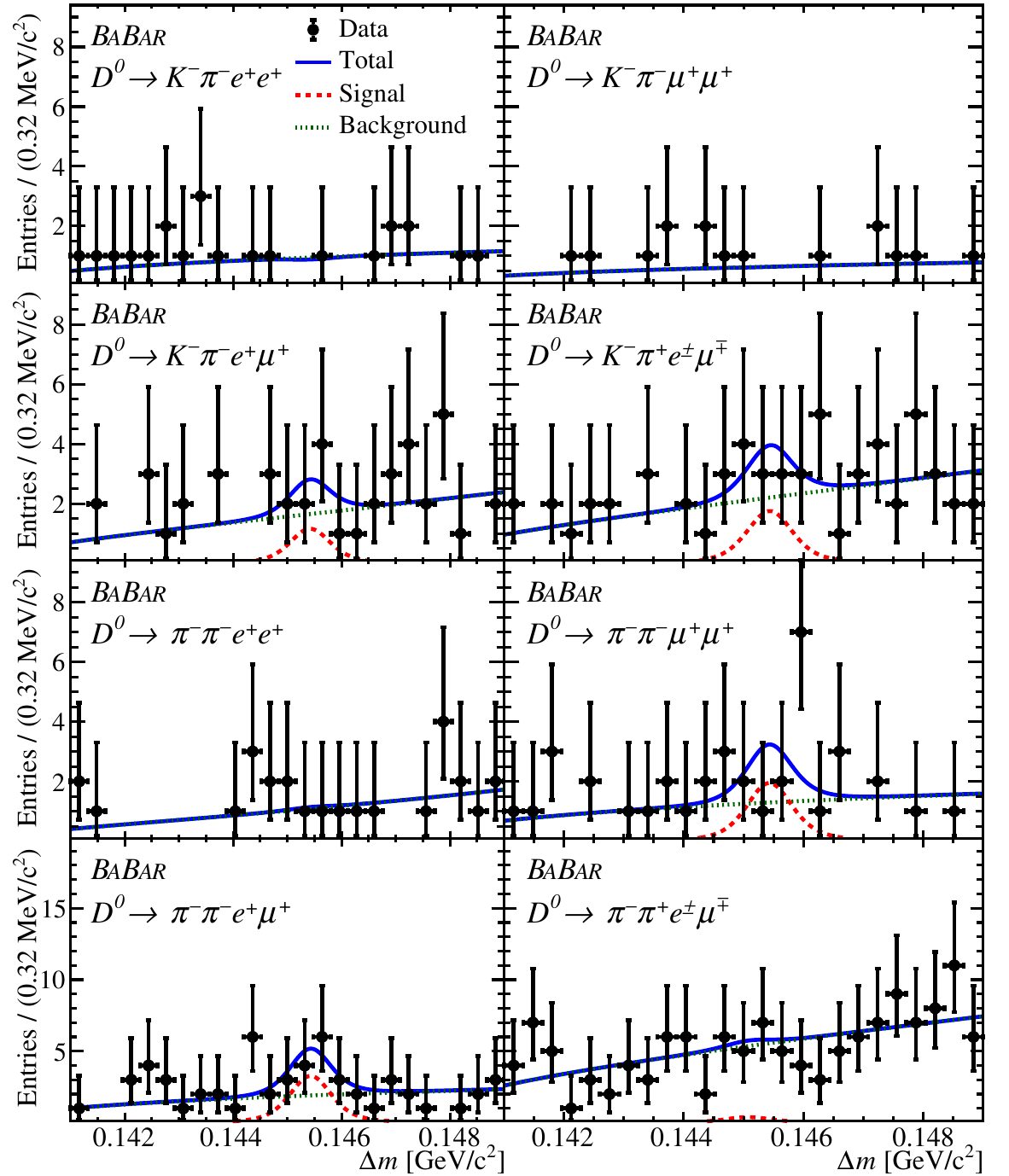}
\end{tabular}
\end{center}
\caption{Projections of the unbinned maximum-likelihood fits to the
  final candidate distributions as a function of \dm\ for the signal
  modes with fewer than two kaons. The solid blue line is the total
  fit, the dashed red line is the signal and the dotted green line is
  the background.}
\label{fig1a}
\end{figure}

\begin{figure}[htbp!]
\begin{center}
  \begin{tabular}{c}
  \includegraphics[width=\columnwidth]{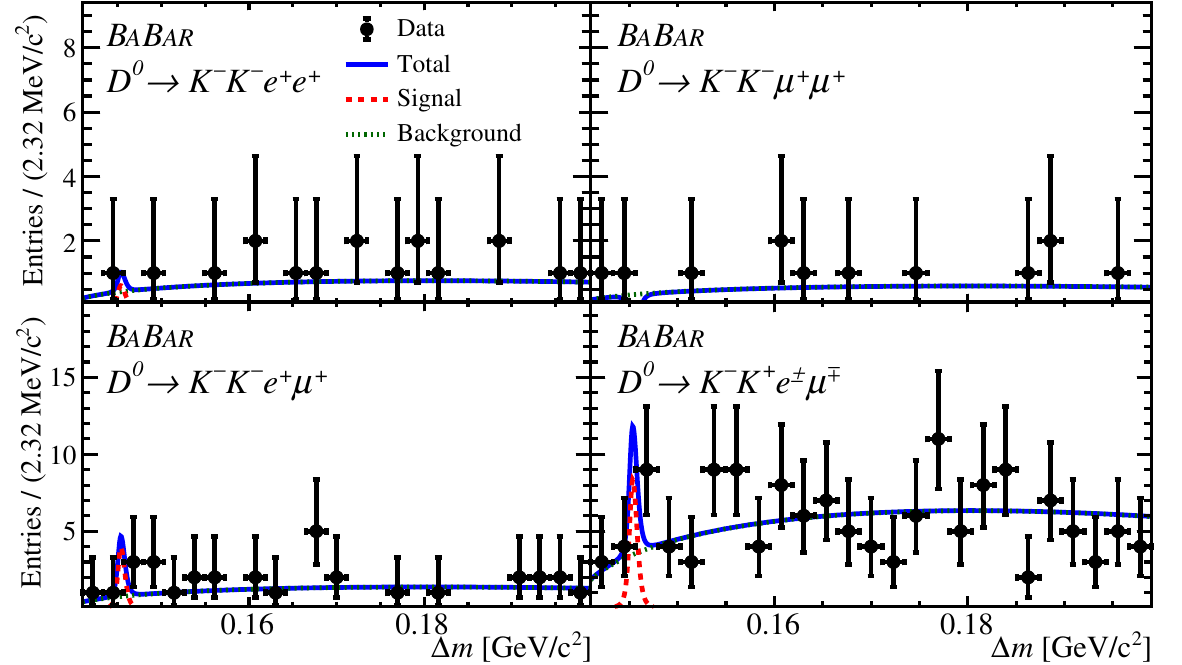}
\end{tabular}
\end{center}
\caption{Projections of the unbinned maximum-likelihood fits to the
  final candidate distributions as a function of \dm\ for the signal
  modes with two kaons. The solid blue line is the total fit, the
  dashed red line is the signal and the dotted green line is the
  background.}
\label{fig2}
\end{figure}

Each signal mode yield \nsig\ is extracted by performing the ML fit
with the single observable \dm\ in the range $0.141<\dm<0.201\gevcc$
for signal modes with two kaons and $0.141<\dm<0.149\gevcc$ for all
other signal modes. The signal PDF is a Cruijff function with
parameters obtained by fitting the signal MC. The background is
modeled with an \argus\ function with an end point that is set to the
same value that is used for the normalization modes. The signal PDF 
parameters and the end point parameter are fixed in the fit. All other
background parameters and the signal and background yields are allowed
to vary. Figures~\ref{fig1a} and~\ref{fig2} show the results of the
fits to the \dm\ distributions for the twelve signal modes.

We test the performance of the ML fit for the normalization modes by
generating ensembles of MC samples from the normalization and
background PDF distributions. The mean numbers of normalization and
background candidates used in the ensembles are taken from the fits to
the data. The numbers of background and normalization mode candidates
are allowed to fluctuate according to a Poisson distribution and all
background and normalization mode PDF parameters are allowed to vary.
No significant biases are observed in fitted yields of the
normalization modes. The same procedure is repeated for the ML fit to
signal modes, with ensembles of MC samples generated from the
background PDF distributions only, assuming a signal yield of
zero. The signal PDF parameters are fixed to the values used for the
fits to the data but the signal yield is allowed to vary. The biases
in the fitted signal yields are less than $\pm0.2$ for all modes, and
these are subtracted from the fitted yields before calculating the
signal branching fractions.

To cross-check the normalization procedure, the signal modes in
Eq.~(\ref{eq:ratio}) are replaced with the decay \DzToKPi, which has a
well-known branching fraction~\cite{PDG2018}. The \DzToKPi\ decay is
selected using the same criteria as used for the \DzToKPiPiPi\ mode,
which is used as the normalization mode for this cross-check. The
\mbox{\DzToKPi} signal yield is \yieldKPi\ with $\epsilon_{\rm sig} =
(\effKPi)\%$.  We determine $\BR(\DzToKPi) = (\BFDzToKPiPiPi)$\%,
where the uncertainties are statistical and systematic,
respectively. This is consistent with the current world average of
$(\BFDzToKPi)$\%~\cite{PDG2018}.  Similar compatibility with the
$\BR(\DzToKPi)$ world average, but with larger uncertainties, is
observed when the normalization mode \DzToKPiPiPi\ in
Eq.~(\ref{eq:ratio}) is replaced with the decay modes \DzToKKPiPi\ or
\DzToPiPiPiPi.

The main sources of systematic uncertainties in the signal yields are
associated with the model parametrizations used in the fits to the
signal modes and backgrounds, the fit biases, and the limited MC and
data sample sizes available for the optimization of the Fisher
discriminants. 

The uncertainties associated with the fit model parametrizations of
the signal modes are estimated by repeating the fits with alternative
PDFs. This involves swapping the Cruijff
  and Crystal Ball functions, using Gaussian functions with different
  asymmetric widths, and changing the number of
  functions used. For the background, the order of the polynomials is
  changed and the \argus\ function is replaced by a second-order
  polynomial. The fits are also performed with the fixed signal
parameters allowed to vary within the statistical uncertainties
obtained from fits to the signal MC samples. The systematic
  uncertainty is taken as half the maximum deviation from the default
  fit.

The systematic uncertainties in the corrections on the fit biases for
the signal yields are taken to be the statistical uncertainties on the
ensembles of fits to the MC samples described above. The systematic
uncertainty due to knowledge of the Fisher discriminant shape is
obtained by varying the value of the selection criterion for the
Fisher discriminant, changing the size of the blinded optimization
region, and retraining the Fisher discriminant using a training sample
with a different set of MC samples. The uncertainty is taken as half
the maximum difference from the yield obtained with the default Fisher
discriminant criterion. Summed together, the total systematic
uncertainties in the signal yield are between 0.4 and 1.9 events,
depending on the mode.

Systematic uncertainties that impact the calculation of the branching
fractions of the signal modes are due to assumptions made about the
distributions of the final-state particles in the signal simulation
modeling, the model parametrizations used in the fits to the
normalization modes, the normalization mode branching fractions,
tracking and PID efficiencies, and luminosity.

Since the decay mechanism of the signal modes is unknown, we vary the
angular distributions of the simulated final-state particles from the
\Dz\ signal decay, where the three angular variables are defined
following the prescription of Ref.~\cite{Aaij:2013cma}. We weight the
reconstruction efficiencies of the phase-space simulation samples as a
function of the angular-variable distributions, trying combinations of
$\sin$, $\cos$, $\sin^2$, and $\cos^2$ functions. Half the maximum
change in the average reconstruction efficiency is assigned as a
systematic uncertainty.

Uncertainties associated with the fit model parametrizations of the
normalization modes are estimated by repeating the fits with
alternative PDFs for the normalization modes and
backgrounds. Uncertainties in the normalization mode branching
fractions are taken from Ref.~\cite{PDG2018}.  We include
reconstruction efficiency uncertainties of 0.8\% per track for the
leptons and 0.7\% for the kaon and
pion~\cite{Allmendinger:2012ch}. For the PID efficiencies, we assign
an uncertainty of 0.7\% per track for electrons, 1.0\% for muons,
0.2\% for pions, and 1.1\% for kaons~\cite{TheBABAR:2013jta}. A
systematic uncertainty of 0.43\% is associated with our knowledge of
the luminosities $\lum_{\rm norm}$ and $\lum_{\rm
  sig}$~\cite{Lees:2013rw}. The total systematic uncertainties in the
signal efficiencies are between 5\% and 19\%, depending on the mode.

We use the frequentist approach of Feldman and
Cousins~\cite{Feldman:1997qc} to determine 90\% C.L. bands that relate
the true values of the branching fractions to the measured signal
yields. When computing the limits, the systematic uncertainties are
combined in quadrature with the statistical uncertainties in the
fitted signal yields.

The signal yields for all the signal modes are compatible with zero.
Table~\ref{tab:hh_results} gives the fitted signal yields,
reconstruction efficiencies, branching fractions with statistical and
systematic uncertainties, and 90\% C.L. branching fraction upper
limits for the signal modes. 

\begin{table*}[htbp!]
\begin{center}
\caption{Summary of fitted signal yields with statistical and
  systematic uncertainties, reconstruction efficiencies, branching
  fractions with statistical and systematic uncertainties, 90\%
  C.L. branching fraction upper limits (U.L.), and the previous
  limits~\cite{PDG2018,Ablikim:2019gvd}. The branching fraction systematic
  uncertainties take into account correlations and cancellations
  between the signal and normalization modes and include the
  uncertainties in the normalization mode branching fractions.}
\begin{tabular}{lrccrrr}   % add some space
\hline
\hline \\ [-1em]
 & & &  & & \multicolumn{2}{c}{\BR\ 90\% U.L. $(\times 10^{-7})$}\\
Decay mode $\Dz\to$    & \nsig\ (candidates) & & $\epsilon_{\rm sig}$ (\%) & \BR $(\times 10^{-7})$ & \babar\ & Previous\\
\hline
$\pim\pim\ep\ep$   &  \obsPiPieeSS   &  & \effPiPieeSS   &
\bfactPiPieeSS   & \ulactPiPieeSS & 1120\\
$\pim\pim\mup\mup$ &  \obsPiPimumuSS &  & \effPiPimumuSS &
\bfactPiPimumuSS & \ulactPiPimumuSS & 290 \\
$\pim\pim\ep\mup$  &  \obsPiPiemuSS  &  & \effPiPiemuSS  &
\bfactPiPiemuSS  & \ulactPiPiemuSS & 790\\
$\pim\pip\epm\mu^{\mp}$ & \obsPiPiemuOS & & \effPiPiemuOS &
\bfactPiPiemuOS  & \ulactPiPiemuOS & 150\\
$\Km\pim\ep\ep$    &  \obsKPieeSS    &  & \effKPieeSS    &
\bfactKPieeSS    & \ulactKPieeSS & 28~\cite{Ablikim:2019gvd}\\
$\Km\pim\mup\mup$  &  \obsKPimumuSS  &  & \effKPimumuSS  &
\bfactKPimumuSS  & \ulactKPimumuSS & 3900\\
$\Km\pim\ep\mup$   &  \obsKPiemuSS   &  & \effKPiemuSS   &
\bfactKPiemuSS   & \ulactKPiemuSS & 2180\\
$\Km\pip\epm\mu^{\mp}$  & \obsKPiemuOS  & & \effKPiemuOS  &
\bfactKPiemuOS   & \ulactKPiemuOS & 5530\\
$\Km\Km\ep\ep$       &  \obsKKeeSS   & & \effKKeeSS   &
\bfactKKeeSS & \ulactKKeeSS & 1520\\
$\Km\Km\mup\mup$     &  \obsKKmumuSS & & \effKKmumuSS &
\bfactKKmumuSS & \ulactKKmumuSS & 940\\
$\Km\Km\ep\mup$      &  \obsKKemuSS  & & \effKKemuSS  &
\bfactKKemuSS & \ulactKKemuSS & 570\\
$\Km\Kp\epm\mu^{\mp}$ &  \obsKKemuOS  & & \effKKemuOS  &
\bfactKKemuOS & \ulactKKemuOS & 1800\\
\hline
\hline
\end{tabular}
\label{tab:hh_results}
\end{center}
\end{table*}

In summary, we report 90\% C.L. upper limits on the branching
fractions for nine lepton-number-violating \mbox{\DzTohhllSS} decays
and three lepton-flavor-violating \mbox{\DzTohhllOS} decays. The
analysis is based on a sample of $\epem$ annihilation data collected
with the \babar\ detector, corresponding to an integrated luminosity
of \totallumi\invfb. The limits are in the range $(\ulactKKmumuSS -
\ulactPiPiemuSS)\times 10^{-7}$ and are between 1 and 3 orders
of magnitude more stringent than previous results.

We are grateful for the excellent luminosity and machine conditions
provided by our \pep\ colleagues, 
and for the substantial dedicated effort from
the computing organizations that support \babar.
The collaborating institutions wish to thank 
SLAC for its support and kind hospitality. 
This work is supported by
DOE
and NSF (USA),
NSERC (Canada),
CEA and
CNRS-IN2P3
(France),
BMBF and DFG
(Germany),
INFN (Italy),
FOM (The Netherlands),
NFR (Norway),
MES (Russia),
MINECO (Spain),
STFC (United Kingdom),
BSF (USA-Israel). 
Individuals have received support from the
Marie Curie EIF (European Union)
and the A.~P.~Sloan Foundation (USA).


\begin{thebibliography}{43}
\bibitem [{con()}]{conjugate}%
  \BibitemOpen
  \href@noop {} {}\bibinfo {note} {Charge conjugation is implied
  throughout.}\BibitemShut {Stop}%
\end{thebibliography}
\end{document}